Closed-form solutions for continuous time random walks on finite chains

Ophir Flomenbom, and Joseph Klafter

*School of Chemistry, Raymond & Beverly Sackler Faculty of Exact Sciences, Tel Aviv University, Ramat Aviv, Tel Aviv 69978, Israel*

Continuous time random walks (CTRWs) on finite arbitrarily inhomogeneous chains are studied. By introducing a technique of counting all possible trajectories, we derive closed-form solutions in Laplace space for the Green's function (propagator) and for the first passage time probability density function (PDF) for nearest neighbor CTRWs in terms of the input waiting time PDFs. These solutions are also the Laplace space solutions of the generalized master equation (GME). Moreover, based on our counting technique, we introduce the adaptor function for expressing higher order propagators (joint PDFs of time-position variables) for CTRWs in terms of Green's functions. Using the derived formulae, an escape problem from a biased chain is considered.

One-dimensional stochastic processes [1-18] are widely used to describe dynamics in biological, chemical, and physical systems [19-33]. Quite ubiquitous among these, are nearest neighbor hopping processes on finite discrete lattices (chains). Yet, closed-form solutions for the dynamics along arbitrarily inhomogeneous chains have been missing. To fill this gap, we consider here a CTRW on a chain of $l$ distinct states governed by state- and direction-dependant waiting time PDFs: $\psi_{n\pm1n}(t) = \omega_{n\pm1n}\varphi_{n\pm1n}(t)$ for transitions between state $n$ to state $n\pm1$, and $\psi_{Nn}(t) = \omega_{Nn}\varphi_{Nn}(t)$ ($N \equiv l+n$) for irreversible trapping from state $n$, with the normalization conditions $\int_0^\infty \varphi_{jn}(t)dt = 1$ $\forall n$, $j = n\pm1, N$ and $\sum_j \omega_{jn} = 1$ $\forall n$ (Fig. 1A). CTRWs with direction-independent exponential waiting time PDFs are Markovian, and those with non-exponential or direction-dependent waiting time PDFs are semi-Markovian [1-3, 14] (see comment below). One can get semi-Markovian processes from Markovian ones by considering branched Markovian processes, where the states along a branch have the same observable value as the backbone-state they have originated from [6], or by averaging over disorder [9].

The dynamics of the CTRW considered here is presented by a stochastic trajectory of $l$ distinct states (Fig. 1B) with the renewal property of having no correlations between waiting times [1-3, 14]. This property relies on two factor: (*i*) the waiting times are drawn from the $\varphi_{jn}(t)$ s randomly and independently of the global time and of the past transitions [1-3, 14], and (*ii*) each state has a distinct observable value [14]. Semi-Markovian processes generate uncorrelated non-exponential waiting times trajectories [2, 14]. Trajectories similar to those shown in Fig. 1B are obtained, for instance, from single molecule measurements [14-32], thus allowing the construction of the waiting time PDFs directly from the trajectory. In many cases, the trajectories consist of two observable states [14-18, 20-21, 25-27, 29-32]. In some of these examples [14, 29, 30a, 31], correlations between waiting times are observed. This non-renewal property of the trajectory can appear when assigning the same observable value for different states of the underlying multi-state chain in a specific way [14]. Here we are interested in renewal processes, although, for some cases, the statistical properties of a non-renewal ones can be also constructed from the results given in this Letter. The Green's function $G_{nm}(t)$, which is the PDF to occupy state $n$ at time $t$ when starting from state $m$ at time $0$ and $(n,m) \in 1,...,l$, obeys for semi-Markovian processes with nearest neighbor transitions, the integro-differential GME,

$$\partial G_{nm}(t)/\partial t = -\int_0^t \left(K_{n+1n}(t-t') + K_{n-1n}(t-t') + K_{Nn}(t-t')\right)G_{nm}(t')dt'$$

$$+ \int_0^t K_{nn-1}(t-t')G_{n-1m}(t')dt' + \int_0^t K_{nn+1}(t-t')G_{n+1m}(t')dt', \qquad n=1,...,l, \qquad (1)$$

with $K_{10}(t) = K_{ll+1}(t) = 0$, and a delta initial condition $(t \to 0)$, $G_{nm}(t) = \delta_{nm}\delta(t)$. Here, $K_{jn}(t)$ is the so-called memory kernel from state $n$ to state $j$, and $K_{Nn}(t)$ is the trapping kernel of state $n$. The relationship between the kernels and the waiting time PDFs is given in Laplace space ($\bar{g}(s) = \int_0^\infty g(t)e^{-st}dt$) by [9, 10, 22],

$\bar{K}_{jn}(s) = \bar{\psi}_{jn}(s)/\bar{\Psi}_n(s)$, where $\bar{\Psi}_n(s) = \left(1 - \sum_i \bar{\psi}_{in}(s)\right)/s$. Two special cases of the GME are obtained for: (*a*) exponential and direction-independent $\varphi_{jn}(t)$s, $\psi_{jn}(t) = a_{jn}e^{-\lambda_n t}$ ($a_{jn}$ is the transition rate from state $n$ to state $j$, and $\lambda_n = \sum_j a_{jn}$), and (*b*) power law and state- and direction-independent $\varphi_{jn}(t)$s, $\psi_{jn}(t) \sim a_{jn}t^{-1-\gamma}$; $0 < \gamma < 1$. For case (*a*), Eq. (1) reduces to the master equation with a sink [11], while for case (*b*), Eq. (1) becomes a fractional master equation with a sink. Note that when taking the continuum limit of case (*b*), one gets a fractional Fokker-Planck equation [13] with a sink.

In this Letter, we derive closed-form solutions in Laplace space for the Green's function that obeys Eq. (1), and for the first passage time PDF, given in terms of the state- and direction-dependent waiting time PDFs [34]. The solutions are obtained by counting all possible trajectories, and are applied here to study an escape problem from a biased chain. Moreover, based on our counting technique, we introduce the adaptor function for expressing higher order propagators for CTRWs in terms of the Green's functions. For the particular case of exponential waiting time PDFs, the factorization of higher order propagators (joint PDFs of time-position variables) into a product of Green's function, known for Markovian processes, is recovered.

To derive a closed-form expression for the Laplace transform of $G_{nm}(t)$, we start from the integral representation of $G_{nm}(t)$, $G_{nm}(t) = \int_0^t W_{nm}(t-\tau)\Psi_n(\tau)d\tau$. Here, $W_{nm}(t)$ is the PDF to reach state *n* exactly at time *t* when starting from state *m* exactly at time 0, and $\Psi_n(t) = \sum_j \Psi_{jn}(t)$, where $\Psi_{jn}(t) = \int_t^\infty \psi_{jn}(\tau)d\tau$, is the sticking probability which is the probability to arrive at state *n* before time *t* and stay there. Using the convolution theorem, we obtain the Laplace space relationship,

$$\overline{G}_{nm}(s) = \overline{W}_{nm}(s)\overline{\Psi}_n(s). \qquad (2)$$

To obtain $\overline{G}_{nm}(s)$ we need to calculate only $\overline{W}_{nm}(s)$, since $\overline{\Psi}_n(s)$ is an input function. We first focus on $\overline{W}_{1l}(s)$. The main idea of our calculation is to express $\overline{W}_{1l}(s)$ as a sum over all possible trajectories (again we use the convenient property of a convolution),

$$\overline{W}_{1l}(s) = \overline{\Gamma}_{1l}(s)\sum_{j=0}^\infty \overline{\alpha}_l(s,j). \qquad (3)$$

Here, $\overline{\Gamma}_{1l}(s)$ is the Laplace transform of the waiting time PDF of the path of direct transitions from state $l$ to state 1. In general, $\overline{\Gamma}_{nm}(s)$ is defined by,

$$\overline{\Gamma}_{nm}(s) = \prod_{i=m}^{n \mp 1} \overline{\psi}_{i \pm 1 i}(s), \ n \neq m; \ \overline{\Gamma}_{mm}(s) = 1, \tag{4}$$

where the upper (lower) sign corresponds to the case $n > m$ ($n < m$). $\overline{\Gamma}_{1l}(s)\overline{\alpha}_l(s,j)$ is the Laplace transform of the waiting time PDF of making $j$ nearest neighbor closed loops when starting at state $l$ and ending at state 1 exactly at time $t$, and the index $l$ stands for the system size. By nearest-neighbor closed loops we mean, for example, a loop made of a transition from state $n$ to state $n+1$ and a back transition from state $n+1$ to state $n$ (not necessarily consecutive transitions). $\overline{\alpha}_l(s,j)$ is obtained by counting all possible continuous trajectories with $j$ nearest neighbor closed loops that started at state $l$ and ended at state 1. Our main objective is to find a general expression for $\overline{\Phi}_l(s)$, which is defined by

$$\overline{\Phi}_l(s) = \left( \sum_{j=0}^{\infty} \overline{\alpha}_l(s,j) \right)^{-1}. \tag{5}$$

We first consider a two-state chain for which $\alpha_2(t,0) = \delta(t)$, and for $j \geq 1$

$\alpha_2(t,j) = \int_0^t \psi_{21}(t-z) \left( \int_0^z \psi_{12}(z-y) \alpha_2(y, j-1) dy \right) dz$. This leads to the recursion relation in

Laplace space $\overline{\alpha}_2(s,j) = (\overline{\psi}_{12}(s)\overline{\psi}_{21}(s))^j = \overline{\alpha}_2^j(s,1)$, and thus to

$\overline{\Phi}_2(s) = 1 - \overline{\alpha}_2(s,1)$.

A recursion relation similar to that of $\overline{\alpha}_2(s,j)$ is found for $\overline{\alpha}_3(s,j)$, for a three-state chain, $\overline{\alpha}_3(s,j) = (\overline{\psi}_{21}(s)\overline{\psi}_{12}(s) + \overline{\psi}_{32}(s)\overline{\psi}_{23}(s))^j = \overline{\alpha}_3^j(s,1)$, which results in

$\overline{\Phi}_3(s) = 1 - \overline{\alpha}_3(s,1)$.

For a four-state chain, the recursion relation for $\overline{\alpha}_4(s,j)$ is more involved. For the first two values of $j$ (= 0, 1), $\overline{\alpha}_4(s,0) = 1$, and $\overline{\alpha}_4(s,1) = \sum_{i=1}^{3} \overline{\psi}_{ii+1}(s)\overline{\psi}_{i+1i}(s)$, as usual. However, $\overline{\alpha}_4(s,2)$ does not obey the relation $\overline{\alpha}_4(s,2) = \overline{\alpha}_4^2(s,1)$, but rather reads $\overline{\alpha}_4(s,2) = \overline{\alpha}_4^2(s,1) - \overline{\beta}_4(s)$, where, $\overline{\beta}_4(s) = \overline{\psi}_{21}(s)\overline{\psi}_{12}(s)\overline{\psi}_{43}(s)\overline{\psi}_{34}(s)$. The term $\overline{\Gamma}_{14}(s)\overline{\beta}_4(s)$ is a broken (discontinuous) trajectory. It is the only broken trajectory in $\overline{\Gamma}_{14}(s)\overline{\alpha}_4^2(s,1)$. This explains the form of $\overline{\alpha}_4(s,2)$. For $j > 1$ we find $\overline{\alpha}_4(s,j) = \overline{\alpha}_4(s,1)\overline{\alpha}_4(s,j-1) - \overline{\beta}_4(s)\overline{\alpha}_4(s,j-2)$, which, after some algebra, leads to,

$$\overline{\Phi}_4(s) = 1 - (\overline{\alpha}_4(s,1) - \overline{\beta}_4(s)).$$

By performing similar calculations for longer chains and using induction, we find that $\overline{\Phi}_l(s)$ is given by,

$$\overline{\Phi}_l(s) = 1 + \sum_{i=1}^{[(l-1)/2, l/2]} (-1)^i \overline{h}_i(s), \quad (6)$$

where $[(l-1)/2, l/2] = l/2$ if $l$ is even and $(l-1)/2$ otherwise ($\overline{\Phi}_l(s)\big|_{l=1} \equiv 1$), and,

$$\overline{h}_i(s) = \prod_{j=1}^{i} \sum_{k_j = k_{j-1}+2}^{l-1-2(i-j)} \overline{\psi}_{k_j k_j+1}(s) \overline{\psi}_{k_j+1 k_j}(s), \quad \text{with } k_0 = -1. \quad (7)$$

Equations (3)-(7) give $\overline{W}_{1l}(s)$. For the case $(m,n) \neq (1,l)$, $\overline{W}_{nm}(s) = \overline{\Gamma}_{nm}(s)\overline{\Phi}_l^{nm}(s)/\overline{\Phi}_l(s)$. $\overline{\Phi}_l^{nm}(s)$ is given by Eq. (6) but for a chain in which the states $n$ and $m$ and those between them are *excluded*, and if after this 'step' a state is left alone at one of the 'edges' of the reduced chain, it is excluded as well (with $\overline{\Phi}_l^{ij}(s) = \overline{\Phi}_l^{ji}(s) = 1$ for $i = 2, j = l-1$). Note that by using $\overline{\Phi}_l^{nm}(s)$, one can define a corresponding $\overline{W}_{1l}^{nm}(s)$. $\overline{W}_{1l}^{nm}(s)$ can then be used to express $\overline{W}_{nm}(s)$. We omit further discussion about this point here. Combining our results, we find that $\overline{G}_{nm}(s)$, which is the Laplace space solution of the GME for any choice of state- and direction-dependent kernels, is given by

$$\overline{G}_{nm}(s) = \overline{\Gamma}_{nm}(s) \frac{\overline{\Phi}_l^{nm}(s)}{\overline{\Phi}_l(s)} \overline{\Psi}_n(s). \tag{8}$$

Equations (4), (6)-(8) are the main results of this Letter. From Eq. (8) one can obtain the first passage time PDF, $F_{Nm}(t)$, which is the PDF of the trapping times to the trap-state $N$ when starting at state $m$, and contains in it the probability to be trapped at state $N$. $\overline{F}_{Nm}(s)$ is obtained when replacing in Eq. (8) $\overline{\Psi}_n(s)$ by $\overline{\psi}_{Nn}(s)$,

$$\overline{F}_{Nm}(s) = \overline{\Gamma}_{nm}(s) \frac{\overline{\Phi}_l^{nm}(s)}{\overline{\Phi}_l(s)} \overline{\psi}_{Nn}(s). \tag{9}$$

We turn now to study, as an example, a biased escape problem from an $l$-state chain with irreversible trapping at each state. The system is characterized by the state- and direction-independent waiting time PDFs, $\psi_{n+1n}(t) = \psi(t)q$, $\psi_{n-1n}(t) = \psi(t)p$, $\psi_{Nn}(t) = \psi(t)w$, and $p + q + w = 1$. To calculate the statistical properties of the process, we need to calculate $\overline{\Phi}_l(s)$. $\overline{\Phi}_l(s)$ is found from Eqs. (6)-(7) to be

$$\overline{\Phi}_l(s) = 2(1/2)^l \overline{Z}_{l+1}(s) / \overline{Z}_1(s), \tag{10}$$

where $\overline{Z}_j(s) = \left(\overline{Q}_+^j(s) - \overline{Q}_-^j(s)\right)$, and $\overline{Q}_\pm(s) = 1 \pm \sqrt{1 - 4pq\overline{\psi}^2(s)}$. Equation (10), when substituted in Eq. (9), generalizes previous results in Ref. [6]. Using Eq. (10), we calculate the mean first passage time, $T$, to reach state 0 when starting at state $l$. We assume that when an irreversible trapping occurs, the process starts again from state $l$ after some characteristic time $t_d$. $T$ is defined by,

$$T = \vartheta \left( \tau_{l+1} + \tau_0 + t_d + \sum_{i=1}^{l} \tau_{l+i} \right), \tag{11}$$

where $\vartheta = 1/\overline{F}_{0l}(0)$ is the average number of events that occurred until an event that terminated at state 0 occurred, and $\tau_j = \int_0^\infty t F_{jl}(t) dt = -[\partial \overline{F}_{jl}(s)/\partial s]_{s=0}$. Substituting Eq. (10) and $\overline{\Phi}_l^{nl}(s) = \overline{\Phi}_{n-1}(s)$ ($n>1$) in Eq. (9), we get in the limit of $q \to 0$,

$$T = t_\psi (cp^{-l} - 1)/w \quad ; \quad t_\psi = \int_0^\infty t\psi(t)dt, \quad (12)$$

where $c = 1 + wt_d/t_\psi$. The dependence of $T$ on the chain length $l$ changes from a *linear* one to an *exponential* one as $w$ changes from $w \to 0$ to $w \to 1$, meaning that by tuning $w$, one can drastically change the system's behavior. This simple model can be used to describe the biological activity of RNA polymerase, which has recently been studied on the single molecule level [33].

Finally, we note that by simple combinations of terms given by Eq. (8) one can obtain the Green's function for a process with a special first event waiting time PDF [2], which is used, for example, when the process has been taking place earlier than the observation started, and the Green's function for a circular (closed-one-dimensional) chain. Moreover, our technique of counting all possible trajectories makes it clear that to get (in Laplace space of appropriate dimensions) higher order propagators for any semi-Markovian hopping process, the properties of only the 'junction waiting time' (Fig. 1B) should be specially considered. Thus, we introduce the 'adaptor' function

$$A_{n'n}(u,s) = \frac{\overline{\psi}_{n'n}(s) - \overline{\psi}_{n'n}(u)}{u - s},$$

for a transition between state $n$ to state $n'$. $A_{n'n}(u,s)$ is the double Laplace transform of the joint backward and forward recurrence times PDF (Fig. 1B). Only for the Markovian case, $A_{n'n}(u,s) = \overline{\Psi}_n(s)\overline{\psi}_{n'n}(u)$, leading to the well-known factorization of higher order propagators into a product of Green's functions. For example, by using the $A_{n\pm 1n}(u,s)$, the double Laplace transform of $G_{knm}(\tau,t) \equiv G_{knm}(t+\tau,t|0)$ ($\tau \to u$ and $t \to s$), which is the PDF to occupy state $k$ at time $\tau + t$ and state $n$ on time $t$

when starting from state $m$ on time $0$, is given for a chain with nearest neighbor transitions by

$$\overline{G}_{knm}(u,s) = \overline{W}_{nm}(s)\left(A_{n-1n}(u,s)\overline{G}_{kn-1}(u) + A_{n+1n}(u,s)\overline{G}_{kn+1}(u)\right). \quad (13)$$

Higher order propagators are useful, for example, in discriminating between different models describing single molecules activities [14, 18, 23, 24]. In particular, $G_{knm}(\tau,t)$ for semi-Markovian process characterized by $\varphi_{jn}(t) \sim t^{-1-\gamma}$; $0 < \gamma < 1$, was given, in the continuum limit, in Ref. [23]. Equation (13), however, can be applied for any choice of the waiting time PDFs, and gives with Eqs. (2)-(10) a detailed characterization of any one-dimensional semi-Markovian hopping process with nearest neighbor transitions.

O. F. thanks Attila Szabo, Zeev Schuss, and Michael Urbakh for stimulating discussions.

[1] W. Feller, *An Introduction to Probability Theory and Its Applications* (Vo. I & Vo. II, John Wiley & Sons, Inc., 1950).

[2] D. R. Cox, *Renewal Theory* (Methuen, London, 1962).

[3] G. H. Weiss, *Aspects and Applications of the Random Walk* (North-Holland, Amsterdam, 1994).

[4] J. Rudnick and G. Gaspari, *Elements of the Random Walk: An introduction for Advanced Students and Researchers*, (Cambridge University Press, 2004).

[5] N. G. van Kampen, *Stochastic Processes in Physics and Chemistry*, revised and enlarged edition (North-Holland, Amsterdam, 1992).

[6] S. Redner, *A Guide to First-Passage Process* (Cambridge University Press, Cambridge, UK, 2001).

[7] N. W. Goel and N. Richter-Dyn, *Stochastic Models in Biology* (Academic Press, New York, 1974).

[8] Z. Schuss, *Theory and Applications of Stochastic Differential Equations*, (John Wiley & Sons, New York, 1980).

[9] J. Klafter and R. Silbey, Phys. Rev. Lett. **44**, 55 (1980).

[10] V. M. Kenkre, E. W. Montroll, and M. F. Shlesinger, J. Stat. Phys. **9**, 45 (1973).

[11] C. Van Den Broeck, M. Bouten, J. Stat. Phys. **45**, 1031 (1986).

[12] M. Boguñá and G. H. Weiss, Physica A **282** 486 (2000); G. H. Weiss, J. Stat. Phys. **24** 587 (1981).

[13] R. Metzler and J. Klafter, Phys. Rep. **339**, 1 (2000).

[14] O. Flomenbom, J. Klafter, and A. Szabo, Biophy. J. **88**, 3780 (2005); O. Flomenbom and J. Klafter, Acta Phys. Pol. B **36**, 1527 (2005); J. Chem. Phys., in press (2005).

[15] J. Cao, J. Chem. Phys. Lett. **327**, 38 (2000); S. Yang and J. Cao, J. Chem. Phys. **117**, 10996 (2002); J. B. Witkoskie and J. Cao, J. Chem. Phys. **121**, 6361 (2004).

[16] G. Margolin and E. Barkai, J. Chem. Phys. **121**, 1566 (2004).

[17] P. Allegrini *et al.*, Phys. Rev. E **68**, 056123 (2003).

[18] J. Wang and P. Wolynes, Phys. Rev. Lett. **74**, 4317 (1995).

[19] A. B. Kolomeisky and M. E. Fisher, J. Chem. Phys. **113**, 10867 (2000).

[20] E. Geva and J. L. Skinner, Chem. Phys. Lett. **288**, 225 (1998); A. M. Berezhkovskii, A. Szabo, G. H. Weiss, J. Phys. Chem. B **104**, 3776 (2000).

[21] E. Barkai, Y. Jung, and R. Silbey, Annu. Rev. Phys. Chem. **55,** 457 (2004).

[22] I. Goychuk, Phys. Rev. E **70**, 016109 (2004).

[23] V. Barsegov and S. Mukamel, J. Phys. Chem. A **108**, 15 (2004).

[24] S. C. Kou and X. Sunney Xie, Phys. Rev. Lett. **93**, 180603 (2004).

[25] B. Hille, *Ion Channels of Excitable Membranes* (Sinauer Associates Inc, USA, 2001).

[26] J. J. Kasianowicz, E. Brandin, D. Branton, and D. W. Deamer, Proc. Natl. Acad. Sci. USA **93**, 13770 (1996).

[27] G. Zumofen, J. Hohlbein, and C. G. Hübner, Phys. Rev. Lett. **93**, 260601 (2004).

[28] H. Yang, *et al*., Science **302**, 262 (2004); W. Min, *et al.,* Phys. Rev. Lett. **94**, 198302 (2005).

[29] X. Zhuang, *et al.*, Science **296**, 1473 (2002); G. Bokinsky, *et al.*, Proc. Natl. Acad. Sci. USA **100**, 9302 (2003);

[30] H. Lu, L. Xun, and X. S. Xie, Science **282**, 1877 (1998); L. Edman, Z. Földes-Papp, S. Wennmalm, and R. Rigler, Chem. Phys. **247**,11 (1999).

[31] O. Flomenbom *et al*., Proc. Natl. Acad. Sci. USA **102**, 2368-2372 (2005); K. Velonia *et al*., Angew. Chem. Int. Ed. **44**, 560-564 (2005).

[32] L. Fleury, A. Zumbusch, M. Orrit, R. Brown, and J. Bernard, J. Lumin. **56**, 15 (1993); I. Chung and M. G. Bawendi, Phys. Rev. B **70**, 165304 (2004).

[33] R. J. Davenport *et al.*, Science **287**, 2497 (2000); K. Adelman *et al.*, Proc. Natl. Acad. Sci. USA. **99**, 13538 (2002); N. R. Forde, *et al.*, Proc. Natl. Acad. Sci. USA. **99**, 11682 (2002).

[34] Note that by utilizing the matrix representation of the GME, one can immediately write down a formal expression for the Green's function. However, a closed form formula for the Green's function in terms of the state- and direction-dependent kernels was not given before using this approach.

**Figure caption**

FIG 1 **A**- Part of an *l*-state semi-Markov chain with nearest neighbor transitions and trapping. The dynamics are governed by the directional waiting time PDFs, $\psi_{n\pm 1n}(t) = \omega_{n\pm 1n}\varphi_{n\pm 1n}(t)$ and $\psi_{Nn}(t) = \omega_{Nn}\varphi_{Nn}(t)$. A way to simulate such a system is to first draw a random number out of a uniform distribution, which determines the direction according to the transition probabilities. Once the direction is set, say from state *n* to state *j*, a time is drawn out of $\varphi_{jn}(t)$, randomly and independently. This Gillespie kind of algorithm produces an uncorrelated waiting times *l*-state trajectory. **B** − Part of a stochastic trajectory, generated from the algorithm above, with *l*=3, $\omega_{21} = \omega_{23} = 1$, $\omega_{12} = \omega_{32} = 1/2$, and $\varphi_{jn} = \lambda^2 t e^{-\lambda t}$. $t'_-$ and $t'_+$ are respectively the backward and forward recurrence times when occupying state 2 at time *t'* and the next transition is to state 1. We call the corresponding waiting time a 'junction waiting time'.

**Figures**

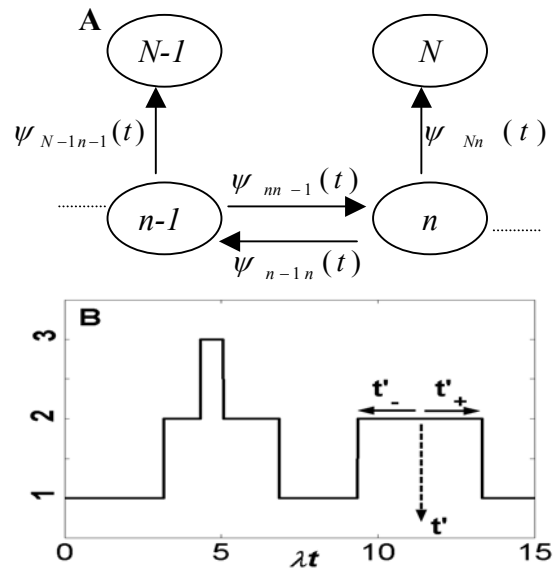

12